\documentclass{article}
\usepackage{arxiv}
\usepackage{booktabs} 
\AtBeginDocument{%
  \providecommand\BibTeX{{%
    Bib\TeX}}}
\usepackage{amsfonts}
\usepackage{algorithmic}
\usepackage{textcomp}
\usepackage{xcolor}
\usepackage{soul}
\def\BibTeX{{\rm B\kern-.05em{\sc i\kern-.025em b}\kern-.08em
    T\kern-.1667em\lower.7ex\hbox{E}\kern-.125emX}}
\usepackage{graphics}
\usepackage{paralist}
\usepackage{color}
\usepackage{balance}
\usepackage{mathtools}
\usepackage[linesnumbered,ruled]{algorithm2e}
\usepackage{bm}
\usepackage{url}
\usepackage{multirow} 
\usepackage{enumitem}
\usepackage{microtype}
\usepackage{array}
\usepackage{hyperref}
\newcolumntype{P}[1]{>{\centering\arraybackslash}p{#1}}
\definecolor{cardinal} {RGB}{196, 30, 58}
\definecolor{lightgrey}{RGB}{150,150,150}
\usepackage[role = cardinal, grid = lightgrey]{credits}

\usepackage[font=small,skip=0pt]{caption}

\usepackage{listings} 
\usepackage{color} 
\definecolor{mygreen}{rgb}{0,0.6,0}
\definecolor{mygray}{rgb}{0.5,0.5,0.5}
\definecolor{mymauve}{rgb}{0.58,0,0.82}
\definecolor{darkgray}{rgb}{.4,.4,.4}
\definecolor{purple}{rgb}{0.65, 0.12, 0.82}
 
\lstdefinelanguage{JavaScript}{
keywords={typeof, new, true, false, catch, function, return, null, catch, switch, var, if, in, while, do, else, case, break},
keywordstyle=\color{blue}\bfseries,
ndkeywords={class, export, boolean, throw, implements, import, this},
ndkeywordstyle=\color{darkgray}\bfseries,
identifierstyle=\color{black},
sensitive=false,
comment=[l]{//},
morecomment=[s]{/*}{*/},
commentstyle=\color{purple}\ttfamily,
stringstyle=\color{red}\ttfamily,
morestring=[b]',
morestring=[b]"
}
 
\graphicspath{{fig/}} 
\newcommand{\sln}{DeFaaS}
\newcommand{\para}[1]{\smallskip\noindent\textbf{#1.}}

\credit{R.K.}{1,1,1,1,0,0,1,1,1,1}
\credit{L.X.}{0,0,0,0,0.5,0.65,1,0,0,1}
\credit{W.S.}{1,0,0,0,1,1,1,0,1,0.6}
\credit{L.C.}{0,0,0,0,0,0,0,0,0,1}

\title{Decentralized FaaS over Multi-Clouds with Blockchain based Management for Supporting Emerging Applications}
\renewcommand{\shorttitle}{\textit{DeFaas}}

\author{ \href{https://orcid.org/0000-0002-6705-6506}{\includegraphics[scale=0.06]{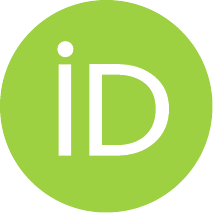}\hspace{1mm}Rabimba Karanjai}\thanks{www.rabimba.me} \\
	Department of Computer Science\\
	University Of Houston\\
	\texttt{rkaranjai@uh.edu} \\
        \And
	Lei Xu \\
 	Department of Computer Science\\
        Kent State University \\
        \And
        Lin Chen\\
        Department of Computer Science\\
        Texas Texh University \\
        \And
        Nour Diallo \\
        Department of Computer Science\\
        University Of Houston \\
        \And
	Weidong Shi \\
 	Department of Computer Science\\
        University Of Houston \\
}

\begin{document}
\maketitle

\begin{abstract}
    Function-as-a-service (FaaS) is a new way to compute in the cloud that further simplifies the user's management burden and allows the user to focus on its core business.
    Most existing FaaS systems are centralized, i.e., the infrastructure is owned and managed by a single cloud service provider.
    This character brings a few limitations.
    For instance, an end user will be bound with a specific cloud service provider and there is a risk of single point of failure.
    %
    In order to mitigate these limitations, a natural idea is to build a decentralized FaaS system.
    In this paper, we propose \sln{}, a novel system that uses blockchain technology and decentralized API management to build a decentralized FaaS system, which enables greater scalability, flexibility, improved security, and reliability.
    %
    %
    Specifically, \sln{} uses a blockchain to create a decentralized registry of available functions (services) and manage their execution. 
    An application interacts with the blockchain to discover and invoke functions securely and transparently.
    %
    In addition to providing a reliable and scalable platform for decentralized FaaS, \sln{} can support other distributed computing scenarios, such as dApps, volunteer computing, and multi-cloud service mesh. 
    Overall, our proposed system represents a significant advance in the field of decentralized computing and has the potential to enable a wide range of exciting new applications and use cases.
\end{abstract}

\section{Introduction}
Cloud computing offers a wide range benefits including high scalability, easy management, and flexibility.
Function-as-a-service (FaaS) is a new way to deliver cloud computation capability and attracts a lot of attention.
Compared with the previous ways of delivering computation services (e.g., virtual machine and container), FaaS further simplifies the management work and allows the user to focus on its own business and the user only needs to pay when a function is executed.
Specifically, a user only needs to provide a set of functions code to the cloud service provider, and then call these functions when needed. 
All related management works (e.g., deployment, execution, and scaling up/down) are shifted to the cloud service provider.

Most existing FaaS systems are centralized, i.e., an FaaS is owned and managed by a single cloud service providers.
This architecture has several limitations:
\begin{inparaenum}[\bfseries (i)]
    \item Binding with a single service provider. When a user selects a cloud platform and submits all its functions code to the corresponding FaaS platform, it is difficult for the user to migrate to another FaaS provider even if the new provider offers better performance/price.
    \item Potential single point of failure. Although cloud is a distributed infrastructure and availability is usually an important part of a typical SLA, it is not uncommon that a cloud data center encounters disruptions for various reasons. 
    \item Incompatibility with emerging decentralized applications. Web3 and dApp are becoming more popular and they also benefit from the FaaS model. However, the nature of Web3 and dApp is not automatically compatible with the single cloud model.
\end{inparaenum}

To mitigate these limitations and harvest the advantages of FaaS, some works have been done on design of multi-cloud FaaS~\cite{https://doi.org/10.48550/arxiv.2209.09367, 10.1145/3472883.3487002}, which allows a user application to utilize FaaS from multiple cloud service providers. 
However all these works require a centralized component to coordinate the interaction between the user application and the FaaS systems. 
Therefore, they only address the binding issue, but cannot contribute much to overcome the other two limitations.

In this work, we propose a transformative and first of its kind decentralized infrastructure for FaaS. The infrastructure is unique in many aspects, including its support for dApp/Web3 applications to take advantage of the cloud resources, its enabling of multi-cloud FaaS for dApp/Web3 developers with a decentralized environment, its complete protocol stack based on open standards, and its adherence to the principle of decentralization without relying on any centralized component.
Furthermore, in this paper, we provide details of a concrete design using Hyperledger Besu and Open FaaS. Although we aim at multi-cloud data centers in this paper, the framework can be extended to encompass managed and volunteer-contributed computing resources, a topic of future research.

This new infrastructure will allow Web3 and dApp developers to take advantage of the scalability and flexibility of cloud computing without sacrificing the decentralized nature of their applications.  To summarize, this paper makes the following main contributions:
\begin{compactitem}
    \item We propose the blockchain-based multi-cloud management architecture to support decentralized function-as-a-service, which is crucial for the deployment of dApps and Web3 applications in multi-cloud environment; 
    \item Details of the critical components design and the way they interact with each other are provided;
    \item We conducted prototyping and preliminary experiments to demonstrate the feasibility and advantages of the proposed system.
\end{compactitem}

\section{Background and Motivation}

\subsection{Decentralized Computing Infrastructure and Cloud}
The future of Internet~\cite{DBLP:journals/corr/abs-2203-00398} will be likely supported by both decentralized ICT infrastructures like computation and storage resources contributed by volunteers~\cite{10.5555/2580126.2580611} and managed ICT infrastructures like general-purpose computing resources provided by the cloud service providers. 

 

Although the market has yet to see widespread adoption of these volunteer-based computing infrastructures (e.g., ~\cite{iexec, golem, sonm, DFinfity_network}), one may argue that they hold great potential to be part of the foundation to support more decentralized infrastructure for today's and future Web applications. On the other hand, there is a rapidly growing adoption of the cloud by the dApp community. For example, 65\% of Ethereum validator nodes are hosted in data centers. Of these nodes, 69\% are using three cloud providers. In some blockchains, like Solana, 95\% of the nodes are hosted in data centers. Some Oracle-oriented dApp platform like API3, recommends their users to run Oracle dApp in the cloud using service like AWS lambda.  


Hosting decentralized applications and computing tasks in multi-cloud environment requires a new control and orchestration layer that effectively harness the resources of cloud data centers and at the same time minimize the risks of centralization. 


\subsection{Advantages of \sln{}}
Deploying dApp and Web3 software to multiple cloud data centers brings many advantages.
Typical benefits of deploying Web3 and dApp software to multi-cloud data centers include avoiding vendor lock-in, improved resilience/availability, increased diversity and better cyber security posture, better performance/scalability, optimized overall cost, and enhanced adherence to the principle of decentralization.

\para{Avoid vendor lock-in} 
Lacking a standardized approach, it often might be impossible for dApp developers to deploy and  migrate dApps across clouds and avoid vendor lock-in. A multi-cloud deployment capable system for dApps gives them flexibility and reliability.

\para{Improved resilience and availability}  
Having dApp and Web3 applications deployed to multi-cloud data centers can increase the availability of the applications to regional users and customers, for instance, Web3 social media applications. 

\para{Increased diversity and cyber security posture}  
Applications hosted in cloud data centers are subject to the whim of the service providers. Although there are service level agreements (SLA) that are designed to provide assurance, SLA for Web3/dApps is still in its infancy. The existing SLA framework for enterprise customers may not be best suited for blockchain and Web3. To ensure continuity of operation, the adoption of multi-cloud for dApps/Web3 is a better approach to reduce exposure to risks. 


\para{Better performance and scalability}  
One provider's data centers may be configured or setup to support certain applications or type of workload better than other providers. Deploying over multi-cloud data centers will give more options to the Web3/dApp developers to optimize performance across geo-distributed data centers based on the unique requirements of the applications. 
Multi-cloud data center deployment can provide more opportunities to optimize performance by globally managing access to different providers' offerings of resources for dApps and Web3 applications. 

\para{Optimized overall cost}  
Different cloud service providers have different billing schemes for services. 
The total cost of a dApp or Web3 application depends on the cloud specific fee models when deployed in different cloud data centers. For running FaaS, the cost depends  not only on the number of requests as well as on other resource measures like memory usage, execution time, network bandwidth consumed. For containerized dApps or Web3 applications, the developers need to consider the differences in terms of fees charged by different cloud providers. 
When dApps/Web3 applications are hosted in different providers' cloud data centers, it is certain that the total cost will be different. There are many subtle pricing differences that dApps/Web3 developers may take into account when balancing the requirements such as budget, diversity, and decentralization. 
Under a holistic optimization scheme decided by the developer team for specific dApp/Web3 use case, the overall cost for deploying dApp or Web3 applications in multi-cloud data centers can be optimized.


\para{Enhanced adherence to the principle of decentralization}  
Considering the fact that high percentages of validators for many public blockchains are hosted in the cloud data centers, stakeholders already propose policy-based validator selection processes like restricting the concentration of validators in certain regions or limiting the number of validators hosted in a single cloud or total validators in all cloud data centers (e.g., ~\cite{beyondstaking}). In similar manner, dApp or Web3 platform can apply policy based approach either directly or indirectly to control the distribution of dApp or Web3 software in multi-cloud data centers. The detailed policies are likely application scenario specific. 


\subsection{System Requirements}
Because of all the advantages of using multi-cloud for Web3 and dApp software, we aim to analyze the design space and provide a general framework to support dApp and Web3 software in the context of a multi-cloud data center environment. 
The key requirements for a general-purpose multi-cloud data center-based environment for the dApp and Web3 applications are summarized as follows:
\begin{inparaenum}[\bfseries (i)]
    \item {\bf Cloud independent APIs.}  Cloud providers often develop specialized services that are mostly compatible only within their own ecosystem.  Using FaaS and serverless computing as example, Gloo~\cite{gloo} is an example framework to support application functions to be deployed in multi-cloud settings. Our endeavor can leverage these efforts to some degree. A fundamental challenge is that these prior or existing multi-cloud efforts focus on centralized setting for enterprise customers. It is often the case that they are not directly applicable to dApps or Web3 where there is no centralized orchestration or coordination entity.
    \item {\bf Adoption of standards.} To support customers, cloud providers have been relying on well-established standards to achieve multi-cloud integration of services. For instance, the OpenAPI Specification (OAS) defines a standard, service provider independent, language-agnostic interface to RESTful APIs~\cite{openapi}. In our case, for invoking dApp functions, instead of using cloud specific API gateways, it is a better option to use cloud agnostic API gateways fully compliant with the OpenAPI specification. 
    \item {\bf Interoperability between cloud services and blockchains.} One main challenge to provide general support of dApp and Web3 deployment in multi-cloud data centers is lack of interoperability between these two ecosystems, enterprise focused cloud ecosystem vs. blockchain centered Web3/dApp ecosystem. To provide some example, cloud and enterprise apply OAuth2 ~\cite{RFC6749,RFC8725} and OpenID~\cite{openid} for identity and authorization management. 
    In the blockchain space, identities are managed using wallet accounts. These two incompatible identity realms can not interoperate with each other directly. It is necessary to develop a bridge solution that can map blockchain identities to the cloud identities. In addition, interoperability is required for access control and authorization process when blockchain users access to the services hosted in multi-cloud data centers. For instance, a blockchain user may access to a dApp using wallet account and Metamask client~\cite{metamask}. If the dApp requires authorization, a solution is needed to provide interoperability.
    \item {\bf Scalability.} The framework should support scalable deployment of dApp and Web3 applications to multi-cloud data centers. Ideally, it should also support use of multi-blockchains. This can be achieved using cross-chain bridges (e.g., ~\cite{ROBINSON2021108488,Herlihy2018AtomicCS, 9756564}).
\end{inparaenum}

\section{Overview of \sln{}}\label{sec-overview}

This section provides the overview of \sln{} and details of each component will be described in Section~\ref{sec-detailed-design}. 

\begin{figure}
	\centering
	\includegraphics[width=3.4in,height=1.9in]{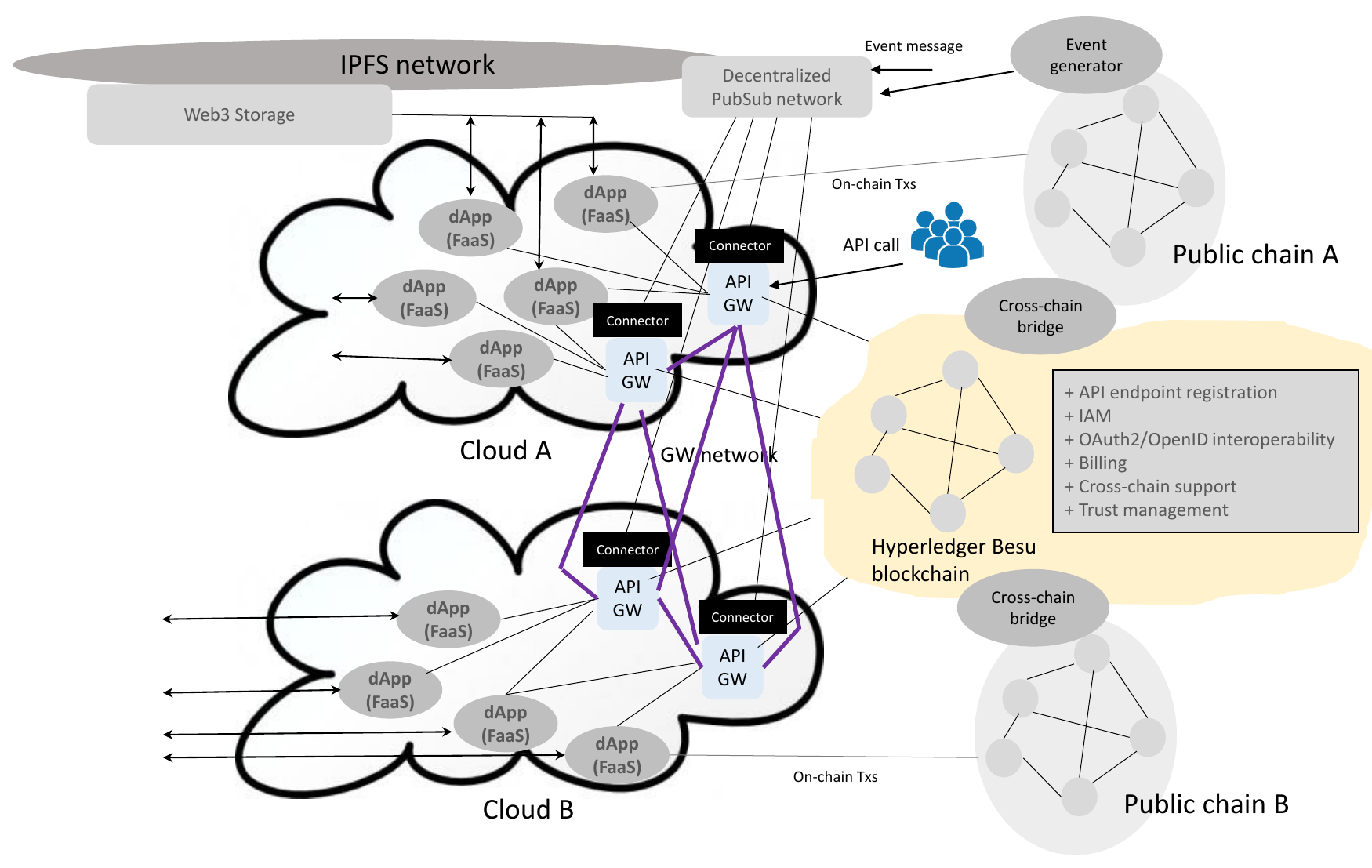}
	\caption{Overview of \sln{}.}   
	\label{fig-overview}
\end{figure}

\subsection{Assumptions} 
This work assumes that dApp and Web3 software is hosted in multi-cloud data centers. The targeted environment focuses on off-chain applications, for instance, off-chain Oracle applications, and light-weighted applications executed off-chain in Web3. This means that services like blockchain as a service (BaaS) or validator as a service (VaaS) are outside the scope of this work. Furthermore, we assume that dApp and Web3 applications can be executed by container-based FaaS. In the future, the framework can be extended to support broad backend services like virtual machine instances. We further make the assumption that the services are invoked using API calls through API gateways. After invoked, execution of a dApp may create blockchain transactions, for instance, update on-chain state/states. However, it is not required that each time, a function is invoked, it must involve any on-chain interaction. One dApp may trigger execution of other dApp functions. A dApp or Web3 application may be triggered by on-chain state update, for instance, run a Web3 function after an on-chain payment is confirmed. For dApp and Web3 applications, IAM is based on blockchain identities. This means that if execution of a function requires authorization, access control is based on blockchain wallet accounts. The environment aims to support dApp and Web3 applications in a multi-blockchain context. We assume the existence of cross-chain bridges for sharing states and exchanging digital assets across blockchains (e.g., ~\cite{ROBINSON2021108488}) like cross-chain billing for services (e.g., ~\cite{Herlihy2018AtomicCS, 9756564}). The cross-chain bridges are assumed to be secure\footnote{Cross-chain bridge security is a research topic of its own, and this work assumes that developers of the bridges have taken necessary steps to protect the bridges and the bridge code has been rigorously audited for security.}. 
There exists trust management scheme for the ecosystem. For instance, stake using tokens may be required for the dApp/Web3 service providers (e.g., API gateway service provider, dApp provider) in the environment. If a service provider misbehaves, its stake may be slashed. Cloud providers are assumed to be somewhat trusted, which means that they will mostly abide by the service level agreements. Cloud providers will not act maliciously and attack the hosted dApp or Web3 applications like launching denial of service attacks. Although this work focuses on multi-cloud setting, the work can be extended in future to a hybrid environment comprising both multi-cloud data centers and user contributed computing resources (a topic of future research). 

\subsection{Overall Architecture of \sln{}}
\figurename~\ref{fig-overview} provides an overview of the system architecture. The framework avoids centralized component, and is fully decentralized in the sense that there is no single entity that can control a major component of the system. Management is achieved using a specific blockchain. At high level, the exact nature of this blockchain, for instance type of consensus protocol, supported smart contract languages,  is irrelevant because one can imagine different implementations to match specific requirements in certain contexts. We assume the existence of a management blockchain. This management blockchain connects to other blockchains using cross-chain bridges for enabling multi-chain support such that dApp and Web3 applications based on different blockchains can be managed by this single management blockchain. Under this framework, a dApp or Web3 function can be invoked by users from multiple blockchains. This management blockchain is also responsible for billing. It accumulates bills for each user. After a period of time, it can send a billing request to the corresponding blockchain to charge the users for the services. For instance, if a user is affiliated with blockchain A and the user has accumulated a total bill of \$100 (say billed in stable coins) in a week. The management chain can use the cross-chain bridge and cross-chain transactions to charge the user on blockchain A. 

The framework supports a network of API gateways hosted in a multi-cloud setting. API gateways are decentralized. The system may require gateway providers to be stake holders. In case that a gateway provider is malicious, its stake can be slashed. The mechanism is similar to how validators are managed in Proof-of-Stake blockchains. The gateway nodes form a distributed mesh network. API calls can be routed among the gateways. 

dApps and Web3 applications are executed on-demand using containerized FaaS (function-as-a-service). 
There are many benefits using such infrastructure. Firstly, container-based FaaS is cloud agnostic, and achieves high portability. Secondly it is secure and agile with the strong isolation of containers. Thirdly, there is no need to plan node capacity and no need to manage server nodes. Fourthly, the system can easily scale out using container based scaling. Fifthly, there is no waste of resources because billing is pay-as-you-go based. 

As illustrated in \figurename~\ref{fig-overview}, the management blockchain offers the main services like API end-point registration, identity management, interoperability support between cloud and blockchain, access control and authorization for the API calls, billing management, policy management, support for monitoring and logging (see details in the next section). At the end, it is worth mentioning that the design does not hold any assumption on the hosting environment of the management blockchain itself. 

\subsection{Prototype Components of \sln{}}
In this specific work, we choose Hyperledger Besu ~\cite{besu} as the management blockchain. It is important to highlight that the overall design is not restricted to Besu or Hyperledger. Besu supports EVM and smart contracts. It can be applied for both public and private blockchain network use cases. For FaaS, 
we use OpenFaaS ~\cite{openfaas}. There are many platforms for FaaS and serverless computing (e.g., ~\cite{whisk,kubeless,fission}). OpenFaaS is selected for demonstration purposes. This means that our framework does not exclude the adoption of other FaaS platforms. OpenFaaS supports containerized FaaS using Kubernetes. It is easy for the developers to deploy event-driven functions and microservices to multi-cloud using OpenFaaS. Applications can be packaged in an OCI-compatible image for deployment. The platform is extensible and customizable. Endpoints are highly scalable with auto-scaling.

\section{Detailed Design of \sln{}}\label{sec-detailed-design}
In this section, we describe the details of the main components of the system. 

\subsection{Decentralized Scheduling and Load Balancing}
Different from the conventional multi-cloud deployment of FaaS, \sln{} avoids any centralized component as a front-end for scheduling API calls.  A user can make API calls to any API gateway. The management blockchain can keep track of the reputation and support trust management of the gateway nodes. Each API gateway needs to be registered with the management blockchain. The system can require a minimal stake to be deposited for the gateway providers. To avoid fragmentation, the system can enforce certain policies like enforcing maximal number of gateways for each cloud, minimal setting of a gateway like types of instances, memory size and the number of vCPUs. 

To dispatch the API calls across cloud data centers, the system applies randomized load balancing (e.g.,~\cite{963420})  to avoid scenarios that all the API calls are sent to a single cloud provider. 
In randomized load balancing, when a gateway receives an API call, depending on the policy setting for the dApp or Web3 application, the gateway randomly routes the call to one of the available cloud service providers.
For instance, the popular power of two heuristic~\cite{mitzenmacher2001power} can achieve good performance with very low overhead. 

A unique characteristic of our design different from all the prior work in randomized load balancing is that scheduling outcome must be publicly verifiable. This is a fundamental difference between distributed system vs. decentralized system. To support public verification, randomization by the gateway nodes must use a random source that can be verified publicly. A simple approach is to use an on-chain random source like block hash. Using block hash like Bitcoin block hash as publicly verifiable random source is a common practice~\cite{Bonneau2015OnBA}. In this work, gateway node applies block hash from a reference chain like Bitcoin or the management blockchain as the random source for randomized load balancing. 

Gateway providers are compensated based on the number of successfully handled API calls. To provide incentives for the gateway providers, payments to the gateways are weighted. This means that if the number of API calls handled by a gateway is skewed from the prediction based on the randomized load balancing and target distribution, its received payment will be reduced. As an example, for a dApp, the system may require all the calls be evenly distributed among three cloud providers. Gateway nodes that cheat the system can be punished, for instance receiving negative payment for the handled API calls that exceed a limit. Scheduling policies are recorded on the management blockchain. Depending on the utility of the use cases, scheduling policies can be configured to achieve proper trade-offs among requirements such as decentralization, cost, performance. For instance, policies prioritizing decentralization may reward distribution of the API calls to multiple clouds. 

To provide scheduling assistance, the system employs Oracle providers who send gateway load data feed to the management blockchain. The Oracle providers estimate gateway load by sending testing API calls and measure performance metrics like the completion time. It is important to mention that the API gateways only query the management blockchain periodically to read other gateways' load status. The gateway load status is cached locally and used for randomized load balancing. There is no on-chain update for viewing gateway load status. 

\subsection{Decentralized Event Distribution}
Event distribution is a critical element for a FaaS platform, including \sln{}. 
It needs particular care in a multi-cloud setting because cloud dependent event mechanism for FaaS (e.g. ~\cite{eventbridge}) is not portable to the multi-cloud environment. Existing approaches to allow functions to consume events from different event sources on multiple clouds (e.g. \cite{triggermesh,knative}) are not designed for Web3 or dApp applications. 
In another word, although they can support cloud agnostic event sources/event consumers, they are not designed for decentralization.
 
In this work, we use GossipSub~\cite{DBLP:journals/corr/abs-2007-02754} to build the event distribution system for dApps and Web3 applications. 
GossipSub is a robust message propagation system adopted by several important blockchain systems including Filecoin and ETH2.0 Network,
and it works in the publish-subscribe mode that relies on a mesh network structure and a score function~\cite{baldoni2007tera} to disseminate messages. 
GossipSub builds on top of floodsub, which is a basic version of the PublishSubscribe system that is used in IPFS and Libp2p. 
It works by sending messages to all nodes in a network, rather than using a more efficient method called CastTree forming~\cite{floodsub}. FloodSub is sometimes referred to as DumbSub or PubSub-Flood.

The mesh network structure allows any node to participate in the message dissemination process, and the score function controls the transmission of messages based on the behavior of individual nodes.
The features of GossipSub makes it a suitable tool for \sln{} to manage events distribution in a decentralized manner across multiple cloud service providers.

Support for function triggers can be implemented as a connector module in OpenFaaS. OpenFaaS is flexible enough to support multiple event trigger sources including third party event sources. In addition, after receiving a triggering event, function can be invoked. GossipSub connector is a module that maps topic based events to the registered functions in OpenFaaS.  

Nodes in the network can connect to each other in two ways: by sending complete messages to sparsely connected nodes called mesh members, or by sharing metadata about the availability of messages. 
Nodes can switch between these two types of connections, known as pruning and grafting, as needed. GossipSub~\cite{vyzovitis2020gossipsub} is designed to be efficient and scalable, with any peer able to participate in the message dissemination process.

In addition to the mesh network and peering structure described previously, GossipSub also includes a system for scoring and limiting the transmission of messages based on the behavior of individual nodes. Nodes that do not reach a certain score threshold~\cite{specspubsubatmasterlibp2pspecs-2022-12-16} will not receive messages. Additionally, nodes can publish messages on topics that they are not subscribed to by sending the message to a small number of randomly selected nodes, called fan-out peers, which will then redistribute the message throughout the network\cite{WhatisPublishSubscribelibp2p-2022-12-16}. The combination of these features makes GossipSub highly efficient, reliable, and resistant to attacks, and allows it to scale effectively~\cite{vyzovitis2020gossipsub}.


\subsection{API Registration and Access Control}
API providers need to register API end-points to the management blockchain. Registration is done through smart contract with a map that associates API end-point with the API provider's account. API end-points can be offered to the public. Alternatively, API end-points can be private (only accessible by certain wallet accounts). Access control policies can be created for each API end-point. The policies are stored as a map from the API end-point and wallet address to a boolean value of true. If the map can be found, it means that the access is permitted.  Note that for public API end-points, there is no need to perform this verification.  Query of access control setting only involves view of on-chain state. 
Listing~\ref{lst-api-reg} shows an example of API registration with permission setting using Solidity.
Since blockchain and smart contract based access control management has been intensively studied in the last few years (e.g.,~\cite{9750531,9057456,accesscontrol,De2020EfficientDA,Almutairi2021}), we skip details to save space and provide references for interested readers ~\cite{9750531}. 

\begin{lstlisting}[caption={Sample Solidity code for API registration with permission setting.}, language=javascript, basicstyle=\tiny, label=lst-api-reg]
    contract APIRegistry{
    // unoptimized example: bool in solidity uses 8 bits 
    struct Permission {
       bool get; 
       bool put;
       bool post;
       bool delete;
    }
    // a map for api path field (string) to a boolean whether the api is public 
    // (no need for access control if in public)
    mapping(string => bool)  isPublicAPI; 
    // a map for api permission based on account address and hash of api path field
    mapping(address => mapping(bytes32 => Permission))  apiPermission; 
    }
\end{lstlisting}

\subsection{OAuth2.0 Support}
API gateways often apply common standard such as OAuth2.0~\cite{RFC6749, RFC6750} and OpenID~\cite{openid} to enable authentication to services and applications.  OAuth2.0~\cite{RFC6749, RFC6750} is the industry-standard protocol for authorization through which an client obtains an access token from an authorization server, to access a protected resource, stored in a resource server.  OAuth 2.0 has received widespread adoption by the cloud providers.  It supports delegation and interoperability, and facilitates access control management. 

To support authorization for API calls, our platform comprises element to enable interoperability between blockchain based authorization and the API gateways using OAuth2.0 access tokens. For this objective, we take advantages of the existing research and prior work that attempt to integrate OAuth2.0 with smart contract and blockchains~\cite{9477026,DBLP:journals/corr/abs-2001-10461}. Our implementation is based on~\cite{DBLP:journals/corr/abs-2001-10461} and OpenFaaS plugin support for authentication.  At high level, the workflow is the following. After verifying a client's permission to access an API end-point, smart contract can generate an ERC token (ERC-721) and OAuth2.0 access token~\cite{RFC7519}. The ERC token is  transferred to the client's wallet account, and the access token is provided to the client. Then, the client can request  access to the API end-point providing the access token. The gateway node can retrieve the corresponding ERC-721 token which is
used for verifying the validity and the ownership of the access token. If the verifications are successful,  the gateway node will route the API call to the targeted function and return the response. In context of OpenFaaS, verification can be implemented as a plugin. Access tokens have expiration time. This reduces the overall performance impact of the described design. 

\subsection{Logging and Billing} 
Our platform enables decentralized logging for the services in multi-cloud data centers. For performance reason, instead of on-chain logging, the decision is to use off-chain logging. It is preferred to use decentralized logging infrastructure. Fortunately, IPFS~\cite{ipfs} has been proposed and experimented in the past as decentralized logging infrastructure\footnote{To protect confidentiality, logs are encrypted when stored over IPFS.}.

The API gateway nodes and API end-point providers may bill the users. For billing, the gateway nodes can keep track of the number of requests from each user, success and response for each request including function execution triggered by the events, and the involved end-point providers. Later, the gateway nodes can bill the users by sending requests and receipts to the management blockchain. If an end-point is offered as a service, the gateway nodes can bill the users on behalf of the end-point providers. 

For performance reasons, billing is not at per request level. The gateway nodes can accumulate statistics, and bill the users when the amount reaches certain threshold or when the time interval reaches a billing cycle (e.g., a month). This means that billing is asynchronous. Such design significantly reduces the overhead
comparing with the alternative that requires on-chain payment transaction for each API call. Users need to provide wallet address and/or make deposit to the management blockchain. For instance, a user may deposit USDC to a designated smart contract first for the next billing cycle\footnote{Note that this step may involve cross-chain transactions.}. The gateway nodes keep a cache of each user's balance. To avoid negative balance, gateway nodes can use a watermark value. When the accumulated bill reaches the watermark, it will send a billing receipt to the management chain. 

On the other hand, in order to prevent a gateway node from overcharging the user, the system requires the gateway node to keep a copy of processed requests. 
Each request includes a nonce and the user creates a digital signature on the request.
Therefore, the gateway node cannot create fake requests and the user cannot deny a submitted request.

\subsection{Trust Management}
Trust management is necessary for a decentralized system. Some of the main entities involved in the system include: {\it cloud providers, API gateway nodes, FaaS end-point providers, blockchain nodes, and users.} As described earlier, cloud providers are somewhat trusted. Note that in our current design, FaaS is mainly for off-chain applications and off-chain processing. The management blockchain (Besu) can be Proof-of-Stake based. Trust management of other entities like  gateway nodes and end-point providers can be stake based, reputation based or a hybrid combining stake and reputation. For instance, staking can be required for the gateway nodes. Entities that deviate from the protocol or act maliciously can be penalized (e.g., slashing). To minimize impact to performance, dispute resolution based process can be applied. The system stores all the API calls and responses in decentralized logs. When there is a dispute, a user can raise a claim against either the involved gateway node or end-point service provider. Then on-chain governance can be used to resolve the dispute. Such mechanism can be found in many blockchain projects like blockchain insurance and on-chain governance of protocol participants. 

\subsection{Supporting Multi-Cloud Service Mesh} 
The concept of service mesh~\cite{li2019service} is an extension of micro service.
With service mesh, a group of micro services are connected to form a processing pipeline.
Existing design and implementation of service mesh adopts the idea of software-defined network (SDN) to separate common functions and service specific functions.
This separation greatly improves micro services from various perspectives, such as efficient management and flexible deployment.
However, existing service mesh architecture does not support multi-cloud. 
This greatly limits the use and adoption of service mesh when an application relies on micro services that belong to multiple cloud service providers.

\sln{} provides all the essential functions to support service mesh cross multiple cloud service providers.
From architecture perspective, the \sln{} management blockchain serves as the coordinator of multiple control planes.
These control planes are managed by different cloud service providers, and each of them manages its own service mesh.
When a request is sent to a service provider, but its corresponding control plane cannot resolve corresponding information, it can query the management blockchain. 
\section{Implementation and Evaluation}

\subsection{Implementations}
The management blockchain is based Hyperledger Besu~\cite{besu}. 
We apply containerized Open FaaS~\cite{openfaas} based on Kubernetes~\cite{kubeless}.  Kubernetes clusters are supported by all the major cloud data centers (e.g., Google cloud, AWS, Azure). The following extensions to the Open FaaS implementation is necessary: a connector that allows OpenFaaS functions to be invoked by the external events where the events are distributed by decentralized IPFS event network; an interoperability plugin for authentication that verifies access tokens from the API clients. Validity of the access tokens are verified against the management blockchain. The implementation is based on~\cite{DBLP:journals/corr/abs-2001-10461} that supports OAuth2 integration using Solidity smart contracts~\cite{sofieoauth}. Support for randomized load balancing is based on modified version of the standard Power of Two Choices. Data sharing between cloud data centers is based on IPFS and Web3 storage. The management blockchain (Besu) supports cross-chain transactions with other public chains. For prototyping purposes, this work makes use of the existing cross-chain contracts (e.g., ~\cite{9569798,sofie}) and bridge components. Among the options for cross-chain bridges, a good choice is SOFIE Interledger bridge~\cite{sofie}.  SOFIE provides bi-directional communications between two blockchains.  It listens for specific events coming from the sender blockchain, and communicates said events to the receiver blockchain, which in turn
acknowledges if the process succeeds. Then the corresponding result is committed in the sender blockchain. Logging of API calls is based on IPFS. 

\subsection{Experiments}
\para{Methodology}
In this section, we will explain how we conducted the measurements in this paper. Firstly, we set up a testbed using storage nodes located in different locations, which were deployed on Amazon AWS and Google Cloud Platform (GCP). Next, we detail the setup used to measure communication operations. Finally, we outline the specific test cases designed to assess the performance.

\para{IPFS testbed}
We have deployed 8 storage nodes for our testbed configuration. These were distributed across 7 geographical locations (6 for GCP) and one local node. The instances of these were deployed in Singapore, Sydney, Frankfurt, Oregon, N. Virginia, Sao Paulo (and for AWS one extra node at Ireland). 
Instances of t2.medium types were used for IPFS nodes (instance configuration as two vCPUs and 4GB memory, Ubuntu LTS 18.04 and IPFS 0.4.18).
The setup is a standard testbed building approach~\cite{hou2017understanding,alcantara2017ginja}. When the data volume of an object is larger than 256KB, IPFS splits the data into multiple blocks. To test the performance of this process we have limited ourselves to 256kb. 
The IPFS environment is tested against application scenarios in \tablename~\ref{tab:trace}. The project(Rsrch) implements a simple dApp for a car rental scenario. 

\subsection{Evaluation}
\begin{figure}
	\centering
 \scriptsize
	\includegraphics[height=1.0in]{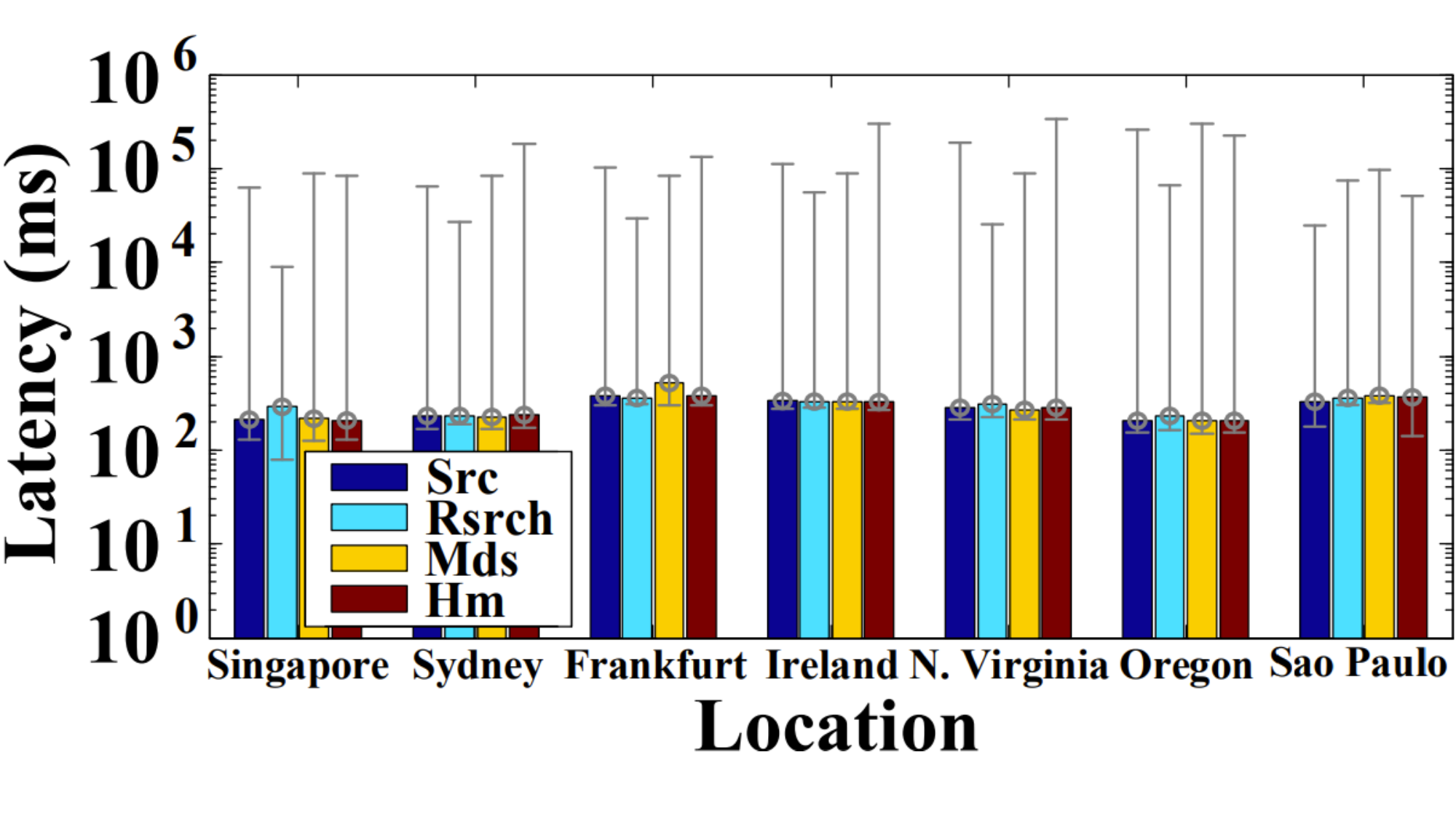}
	\caption{Latency ranges for event messages at different locations.}   
	\label{fig-latency}
\end{figure}

\para{Multi-cloud event dissemination}
Multi-cloud event distribution performance is evaluated. We first present the latency with different applications from remote nodes in \figurename~\ref{fig-latency}. For small messages (\figurename~\ref{fig-latency}), the latency varies significantly across different remote nodes. In particular, the maximum latency can be much higher than the median value. 

To evaluate the performance for a given message size on each node, we run the experiment 30 times for selected applications. For \figurename~\ref{fig-overview} we see that when communicating between different clouds, what affects the event latency more is the different geographical locations than different cloud providers. Since most of the cloud providers are already very optimized, inter-cloud communications have a lesser impact than actual physical servers being in different geographical regions than our testbed and also from the different servers sending the different-sized packets to IPFS. Note that the differences between applications are small and negligible. 

Out current work is to integrate with a new decentralized network for disseminating information on top of the IPFS, called SmartPubSub ~\cite{https://doi.org/10.48550/arxiv.2207.06369}. SmartPubSub leverages  ScoutSubs, a completely decentralized protocol for event message distribution and a fast delivery protocol centered around the publisher. Based on the experiments with a network of containerized SmartPubSub nodes (in one region), ~\cite{https://doi.org/10.48550/arxiv.2207.06369}, the average event latency under normal situation is between 200 - 250ms. 

\para{Effect of IPFS read operations}
We measure performance of the downloading operations across clouds (data saved to the IPFS and retrieved using read operations by another node). Latency is determined by the length of time it takes for an operation to complete, and throughput is calculated by dividing the amount of data processed by the latency of the operation. We observe that the locations of the remote nodes do not have a major impact on the read latency. This is likely because the experimented applications often perform small read requests, which can result in high variance in read latency. When a node performs remote read operations, the resolving and downloading operations required by the local node significantly affect the I/O performance.

\begin{table}[]
\centering
\scriptsize
\caption{\label{tab:trace}Test application names \& descriptions.}
\begin{tabular}{@{}|l|l|@{}}
\toprule
Name & Modules using IPFS for communication \\ \midrule
User       & User home directories                \\ 
Web        & Web/SQL server                       \\ 
Proj       & Project directories                  \\ 
Ts         & Terminal server                      \\ 
Src        & Source control (Subversion)          \\ 
Rsrch      & Research projects (Carpooling dApp)  \\ 
Mds        & Media server (kodi)                  \\ 
Hm         & Hardware monitoring (hwmon)          \\ \bottomrule
\end{tabular}
\end{table}

\begin{figure}
	\centering
 \scriptsize
	\includegraphics[height=1.0in]{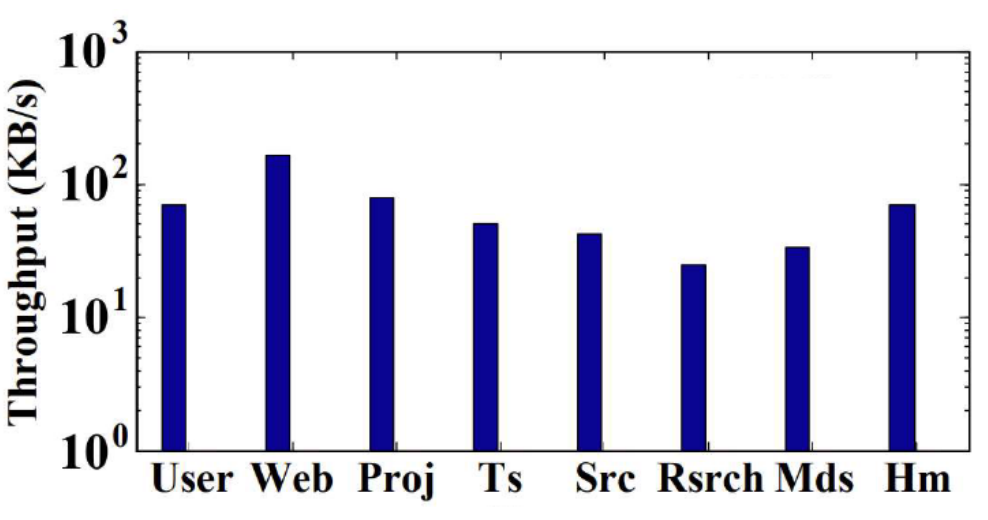}
	\caption{Throughput of read operations over IPFS.}   
	\label{fig-Throughput}
\end{figure}


\para{Randomized load balancing} 
To evaluate the performance of randomized load balancing, we have developed a simulator for experiments. 
The simulator models the API gateways and routing of API calls to multi-cloud data centers. 
In the current version, API calls take the same unit time to complete, which can be easily extended to random completion time.
The simulator allows us to explore load balancing performance of different heuristics: 
\begin{inparaenum}[(i)]
    \item a default design based on the standard power of two choices; 
    \item a simplified heuristic called choice of two; and
    \item a baseline of no randomized load balancing. 
\end{inparaenum}
In the default design, when an API gateway of a cloud data center receives a call, it will randomly pick another cloud data center using on-chain random source. If the receiving data center is busier than the randomly picked data center, then the call will be forwarded to the randomly picked data center. In the choice of two heuristic,  when an API gateway of a cloud data center receives a call, it will evaluate whether the receiving cloud has been overloaded (queued calls exceeding its configured capability). If the receiving cloud is overloaded, it will forward the call to a randomly picked data center. In the baseline, there is no randomized load balancing. Due to space limit, here we report the results assuming that calls should be evenly distributed to the data centers. Further, when a call is forwarded, gateway node will be randomly selected. 

\begin{table}
\centering
\scriptsize
\caption{Performance of randomized load balancing (average queuing times in different scenarios: assume api call processed with one time unit, see text for explanation)}
\label{tab:random}
\scriptsize
\begin{tabular}{p{5.2cm}|p{0.6cm}|p{0.6cm}|p{0.6cm}}
\hline \hline
& 6 data centers  & 8 data centers & 10 data centers \\ \hline
Default implementation (randomly distributed calls) & 2.515 &  2.543 & 2.680  \\ \hline
Default implementation (calls initially sent to one data center) & 2.574 & 2.574 & 2.826 \\ \hline
Choice of two (randomly distributed calls) & 2.915 & 3.071 & 3.302 \\ \hline
Choice of two (calls initially sent to one data center)  & 3.029 & 3.199 & 3,657 \\ \hline
Randomly distributed calls (no load balancing) & 12.397 & 12.353 & 10.690 \\
\hline \hline
\end{tabular}
\end{table}

We have simulated three settings, i.e., 6, 8, and 10 data centers. Results are averaged based on five runs of the simulation. Each run simulates 10,000 API calls. \tablename~\ref{tab:random} shows the average API call queue times for different experiment scenarios. According to the results, the default implementation of randomized load balancing achieves the best performance in terms of average queue times comparing with the other options. The second option (choice of two) only performs slightly worse. The average queue times are the worst without randomized load balancing.  Although the second option has slightly longer average queue times, it may still be a good candidate in the actual deployment because of its simplicity. In the second option, a gateway node is not required to track or query any state status of other data centers. The support for a lightweight implementation (no need to keep states of other data centers) may overcome its slight disadvantage in performance. Further, the heuristic and simulator can be modified to use multi-dimension utility that combines cost, decentralization and performance. 

\section{Related Work}
The related work can be categorized into the following areas. 

\para{Multi-cloud FaaS} 
There exist efforts like~\cite{https://doi.org/10.48550/arxiv.2209.09367, 10.1145/3472883.3487002} to enable multi-cloud FaaS. Since our design is described in the context of Open FaaS, it is worth mentioning the existing effort on supporting distributed Open FaaS~\cite{Vieira2019DisOpenFaaSAD}. The objectives of these efforts are fundamentally different from ours as they mainly target enterprise customers and enterprise application settings. For this reason, they often involve centralized component for scheduling, orchestration, synchronization or monitoring purposes, for instance, the controller in~\cite{10.1145/3472883.3487002}. In addition, they do not need to offer integration or interopreability with blockchains for instance support for authorization and authentication based on blockchain wallet accounts. The trust model and assumptions are also drastically different. Furthermore, a major distinguishing characteristic is that our design and protocol  primarily rely on decentralized building components for almost all the major functions like event triggering, state synchronization, management, scheduling, authorization and authentication. 

\para{dApp/Web3 offering by cloud providers} 
As described earlier, to capitalize the rapid growth of Web3, cloud providers are rushing to this space by offering cloud based resources to Web3 and dApp developers. To certain extent, our efforts are both complementary to and different from the cloud providers' efforts. Regarding the differences, our efforts mainly target at the areas of multi-cloud support, decentralized management layer, blockchain integration and interoperability. These elements are missing in Web3 strategy of a single cloud provider. For instance, it is not at a single cloud provider's interest to enable multi-cloud support of FaaS for dApp and Web3. Decentralized management layer above a single cloud's realm is also out of the reach of a single cloud provider. In another word, our design fills a gap that is both outside the interest as well as outside the reach of a single cloud provider. 

\para{Cloud-based dApps and Web3 applications}  
Web3 and dApp projects have been applying cloud services for many advantages such as low cost, on-demand resources, out-sourced management. There exist efforts to facilitate integration of cloud based dApps with blockchains like API3~\cite{api3} and Chainlink~\cite{10.1007/978-3-662-63958-0_10}. Our work  fundamentally distinguishes from them in both the overall goals and implementations. Firstly, API3 and Chainlink mainly focus on Oracle applications, dApps that provide Oracle data feeds to the blockchains. Our design is not restricted to Oracle applications. From the description, one can easily tell that our framework is a comprehensive approach for decentralized FaaS, which is much broader than what exist today. Secondly, we provide a unified and blockchain based management layer that spans multi-cloud data centers. Such component is missing in the related work. For instance, API3 relies on cloud provider's API gateway service. In addition, it does not support coordination across multi-cloud data centers. In contrast, our framework enables decentralized FaaS across multi-cloud with a network of API gateways. Most of the functions and components supported by our framework such as decentralized API scheduling, decentralized event triggering, and decentralized state synchronization, etc are missing in the related work. Partially the reason is that they focus on a much narrower objective, leveraging  the cloud resources for off-chain Oracle sources, whereas our goal is to enable fully decentralized FaaS. The dramatic differences in the objective lead to completely different approaches in design. 

\para{Volunteer computing-based support for dApp/Web3 applications} 
Like mentioned earlier, there exist  marketplaces and frameworks to enable sharing and exchange computing resources contributed by volunteers (e.g., ~\cite{iexec,golem, sonm, DFinfity_network}). In almost all the cases, these projects focus on tokenization of computing resources and electronic marketplaces for trading the tokenized computing resources. Although our framework can be easily extended to include a decentralized marketplace component, this is not the main focus of the described work. Furthermore, our design focuses on decentralized FaaS deployed over multi-cloud data centers instead of resources contributed from anonymous users or peers. Our focus on decentralized management and governance of decentralized FaaS is not addressed in these volunteer computing based projects. However, it is important to point out that it is plausible to extend our framework to cover resources from non-cloud providers like ICT resources from volunteers. 

\section{Conclusions}
FaaS is a new computation paradigm, and offers high flexible and scalable computation capability.
While it is feasible for an dApp/Web3 application to take advantage of the existing FaaS, it greatly impacts the decentralization character as most FaaS is owned and managed by a single cloud service provider.
We propose \sln{} in this work, which utilizes blockchain technology to coordinate FaaS systems of multiple cloud service providers.

\balance
\bibliographystyle{ACM-Reference-Format}
\bibliography{ref}


\begin{thebibliography}{54}


\ifx \showCODEN    \undefined \def \showCODEN     #1{\unskip}     \fi
\ifx \showDOI      \undefined \def \showDOI       #1{#1}\fi
\ifx \showISBNx    \undefined \def \showISBNx     #1{\unskip}     \fi
\ifx \showISBNxiii \undefined \def \showISBNxiii  #1{\unskip}     \fi
\ifx \showISSN     \undefined \def \showISSN      #1{\unskip}     \fi
\ifx \showLCCN     \undefined \def \showLCCN      #1{\unskip}     \fi
\ifx \shownote     \undefined \def \shownote      #1{#1}          \fi
\ifx \showarticletitle \undefined \def \showarticletitle #1{#1}   \fi
\ifx \showURL      \undefined \def \showURL       {\relax}        \fi
\providecommand\bibfield[2]{#2}
\providecommand\bibinfo[2]{#2}
\providecommand\natexlab[1]{#1}
\providecommand\showeprint[2][]{arXiv:#2}

\bibitem[Agostinho et~al\mbox{.}(2022)]%
        {https://doi.org/10.48550/arxiv.2207.06369}
\bibfield{author}{\bibinfo{person}{Pedro Agostinho}, \bibinfo{person}{David Dias}, {and} \bibinfo{person}{Luís Veiga}.} \bibinfo{year}{2022}\natexlab{}.
\newblock \bibinfo{title}{SmartPubSub: Content-based Pub-Sub on IPFS}.
\newblock
\newblock
\urldef\tempurl%
\url{https://doi.org/10.48550/ARXIV.2207.06369}
\showDOI{\tempurl}


\bibitem[Alc{\^a}ntara et~al\mbox{.}(2017)]%
        {alcantara2017ginja}
\bibfield{author}{\bibinfo{person}{Joel Alc{\^a}ntara}, \bibinfo{person}{Tiago Oliveira}, {and} \bibinfo{person}{Alysson Bessani}.} \bibinfo{year}{2017}\natexlab{}.
\newblock \showarticletitle{Ginja: One-dollar cloud-based disaster recovery for databases}. In \bibinfo{booktitle}{\emph{Proceedings of the 18th ACM/IFIP/USENIX Middleware Conference}}. \bibinfo{pages}{248--260}.
\newblock


\bibitem[Almutairi et~al\mbox{.}(2021)]%
        {Almutairi2021}
\bibfield{author}{\bibinfo{person}{Suzan Almutairi}, \bibinfo{person}{Nusaybah Alghanmi}, {and} \bibinfo{person}{Muhammad~Mostafa Monowar}.} \bibinfo{year}{2021}\natexlab{}.
\newblock \showarticletitle{Survey of Centralized and Decentralized Access Control Models in Cloud Computing}.
\newblock \bibinfo{journal}{\emph{International Journal of Advanced Computer Science and Applications}} \bibinfo{volume}{12}, \bibinfo{number}{2} (\bibinfo{year}{2021}).
\newblock
\urldef\tempurl%
\url{https://doi.org/10.14569/IJACSA.2021.0120243}
\showDOI{\tempurl}


\bibitem[Amazon({[n.\,d.]})]%
        {eventbridge}
\bibfield{author}{\bibinfo{person}{Amazon}.} \bibinfo{year}{[n.\,d.]}\natexlab{}.
\newblock \bibinfo{title}{EventBridge}.
\newblock \bibinfo{howpublished}{\url{https://aws.amazon.com/eventbridge/}}.
\newblock
\newblock
\shownote{Last accessed on 9/28/2020}.


\bibitem[Baarzi et~al\mbox{.}(2021)]%
        {10.1145/3472883.3487002}
\bibfield{author}{\bibinfo{person}{Ataollah~Fatahi Baarzi}, \bibinfo{person}{George Kesidis}, \bibinfo{person}{Carlee Joe-Wong}, {and} \bibinfo{person}{Mohammad Shahrad}.} \bibinfo{year}{2021}\natexlab{}.
\newblock \showarticletitle{On Merits and Viability of Multi-Cloud Serverless}. In \bibinfo{booktitle}{\emph{Proceedings of the ACM Symposium on Cloud Computing}} (Seattle, WA, USA) \emph{(\bibinfo{series}{SoCC '21})}. \bibinfo{publisher}{Association for Computing Machinery}, \bibinfo{address}{New York, NY, USA}, \bibinfo{pages}{600–608}.
\newblock
\showISBNx{9781450386388}
\urldef\tempurl%
\url{https://doi.org/10.1145/3472883.3487002}
\showDOI{\tempurl}


\bibitem[Baldoni et~al\mbox{.}(2007)]%
        {baldoni2007tera}
\bibfield{author}{\bibinfo{person}{Roberto Baldoni}, \bibinfo{person}{Roberto Beraldi}, \bibinfo{person}{Vivien Quema}, \bibinfo{person}{Leonardo Querzoni}, {and} \bibinfo{person}{Sara Tucci-Piergiovanni}.} \bibinfo{year}{2007}\natexlab{}.
\newblock \showarticletitle{TERA: topic-based event routing for peer-to-peer architectures}. In \bibinfo{booktitle}{\emph{Proceedings of the 2007 inaugural international conference on Distributed event-based systems}}. \bibinfo{pages}{2--13}.
\newblock


\bibitem[Bambacht and Pouwelse(2022)]%
        {DBLP:journals/corr/abs-2203-00398}
\bibfield{author}{\bibinfo{person}{Joost Bambacht} {and} \bibinfo{person}{Johan Pouwelse}.} \bibinfo{year}{2022}\natexlab{}.
\newblock \showarticletitle{Web3: {A} Decentralized Societal Infrastructure for Identity, Trust, Money, and Data}.
\newblock \bibinfo{journal}{\emph{CoRR}}  \bibinfo{volume}{abs/2203.00398} (\bibinfo{year}{2022}).
\newblock
\urldef\tempurl%
\url{https://doi.org/10.48550/arXiv.2203.00398}
\showDOI{\tempurl}
\showeprint[arXiv]{2203.00398}


\bibitem[Benet(2014)]%
        {ipfs}
\bibfield{author}{\bibinfo{person}{Juan Benet}.} \bibinfo{year}{2014}\natexlab{}.
\newblock \bibinfo{title}{IPFS - Content Addressed, Versioned, P2P File System}.
\newblock
\newblock
\urldef\tempurl%
\url{https://doi.org/10.48550/ARXIV.1407.3561}
\showDOI{\tempurl}


\bibitem[Besu({[n.\,d.]})]%
        {besu}
\bibfield{author}{\bibinfo{person}{Besu}.} \bibinfo{year}{[n.\,d.]}\natexlab{}.
\newblock \bibinfo{title}{Hyperledger Besu}.
\newblock \bibinfo{howpublished}{\url{https://besu.hyperledger.org/}}.
\newblock
\newblock
\shownote{Accessed: 04/20/2021}.


\bibitem[Bonneau et~al\mbox{.}(2015)]%
        {Bonneau2015OnBA}
\bibfield{author}{\bibinfo{person}{Joseph Bonneau}, \bibinfo{person}{Jeremy Clark}, {and} \bibinfo{person}{Steven Goldfeder}.} \bibinfo{year}{2015}\natexlab{}.
\newblock \showarticletitle{On Bitcoin as a public randomness source}.
\newblock \bibinfo{journal}{\emph{IACR Cryptol. ePrint Arch.}}  \bibinfo{volume}{2015} (\bibinfo{year}{2015}), \bibinfo{pages}{1015}.
\newblock


\bibitem[Burak~Benligiray({[n.\,d.]})]%
        {api3}
\bibfield{author}{\bibinfo{person}{Heikki~V¨anttinen Burak~Benligiray, Saˇsa~Mili´c}.} \bibinfo{year}{[n.\,d.]}\natexlab{}.
\newblock \bibinfo{title}{Decentralized APIs for Web 3.0}.
\newblock \bibinfo{howpublished}{\url{https://api3.org}}.
\newblock


\bibitem[Cha et~al\mbox{.}(2021)]%
        {9477026}
\bibfield{author}{\bibinfo{person}{Shi-Cho Cha}, \bibinfo{person}{Chu-Lin Chang}, \bibinfo{person}{Yang Xiang}, \bibinfo{person}{Zi-Jia Huang}, {and} \bibinfo{person}{Kuo-Hui Yeh}.} \bibinfo{year}{2021}\natexlab{}.
\newblock \showarticletitle{Enhancing OAuth with Blockchain Technologies for Data Portability}.
\newblock \bibinfo{journal}{\emph{IEEE Transactions on Cloud Computing}} (\bibinfo{year}{2021}), \bibinfo{pages}{1--1}.
\newblock
\urldef\tempurl%
\url{https://doi.org/10.1109/TCC.2021.3094846}
\showDOI{\tempurl}


\bibitem[De and Ruj(2020)]%
        {De2020EfficientDA}
\bibfield{author}{\bibinfo{person}{Sourya~Joyee De} {and} \bibinfo{person}{Sushmita Ruj}.} \bibinfo{year}{2020}\natexlab{}.
\newblock \showarticletitle{Efficient Decentralized Attribute Based Access Control for Mobile Clouds}.
\newblock \bibinfo{journal}{\emph{IEEE Transactions on Cloud Computing}}  \bibinfo{volume}{8} (\bibinfo{year}{2020}), \bibinfo{pages}{124--137}.
\newblock


\bibitem[Fission({[n.\,d.]})]%
        {fission}
\bibfield{author}{\bibinfo{person}{Fission}.} \bibinfo{year}{[n.\,d.]}\natexlab{}.
\newblock \bibinfo{title}{Fast serverless functions for kubernetes}.
\newblock \bibinfo{howpublished}{\url{https://github.com/fission/fission}}.
\newblock
\newblock
\shownote{Accessed: 01/20/2021}.


\bibitem[Fotiou et~al\mbox{.}(2020)]%
        {DBLP:journals/corr/abs-2001-10461}
\bibfield{author}{\bibinfo{person}{Nikos Fotiou}, \bibinfo{person}{Iakovos Pittaras}, \bibinfo{person}{Vasilios~A. Siris}, \bibinfo{person}{Spyros Voulgaris}, {and} \bibinfo{person}{George~C. Polyzos}.} \bibinfo{year}{2020}\natexlab{}.
\newblock \showarticletitle{OAuth 2.0 authorization using blockchain-based tokens}.
\newblock \bibinfo{journal}{\emph{NDSS Workshop on Decentralized IoT Systems and Security (DISS)}}  \bibinfo{volume}{abs/2001.10461} (\bibinfo{year}{2020}).
\newblock
\showeprint[arXiv]{2001.10461}
\urldef\tempurl%
\url{https://arxiv.org/abs/2001.10461}
\showURL{%
\tempurl}


\bibitem[Github({[n.\,d.]})]%
        {sofie}
\bibfield{author}{\bibinfo{person}{Github}.} \bibinfo{year}{[n.\,d.]}\natexlab{}.
\newblock \bibinfo{title}{Sofie interledger repository}.
\newblock \bibinfo{howpublished}{\url{https://github:com/SOFIE-project/Interledger}}.
\newblock


\bibitem[Gloo({[n.\,d.]})]%
        {gloo}
\bibfield{author}{\bibinfo{person}{Gloo}.} \bibinfo{year}{[n.\,d.]}\natexlab{}.
\newblock \bibinfo{title}{An Envoy-Powered API Gateway}.
\newblock \bibinfo{howpublished}{\url{https://docs.solo.io/gloo-edge/latest/}}.
\newblock


\bibitem[Hanke et~al\mbox{.}(2018)]%
        {DFinfity_network}
\bibfield{author}{\bibinfo{person}{Timo Hanke}, \bibinfo{person}{Mahnush Movahedi}, {and} \bibinfo{person}{Dominic Williams}.} \bibinfo{year}{2018}\natexlab{}.
\newblock \bibinfo{title}{DFINITY Technology Overview Series, Consensus System}.
\newblock
\newblock
\urldef\tempurl%
\url{https://doi.org/10.48550/ARXIV.1805.04548}
\showDOI{\tempurl}


\bibitem[Hardt(2012)]%
        {RFC6749}
\bibfield{author}{\bibinfo{person}{D. Hardt}.} \bibinfo{year}{2012}\natexlab{}.
\newblock \bibinfo{booktitle}{\emph{The OAuth 2.0 Authorization Framework}}.
\newblock \bibinfo{type}{RFC} 6749. \bibinfo{institution}{RFC Editor}.
\newblock
\showISSN{2070-1721}
\urldef\tempurl%
\url{http://www.rfc-editor.org/rfc/rfc6749.txt}
\showURL{%
\tempurl}
\newblock
\shownote{\url{http://www.rfc-editor.org/rfc/rfc6749.txt}}.


\bibitem[Herlihy(2018)]%
        {Herlihy2018AtomicCS}
\bibfield{author}{\bibinfo{person}{Maurice Herlihy}.} \bibinfo{year}{2018}\natexlab{}.
\newblock \showarticletitle{Atomic Cross-Chain Swaps}.
\newblock \bibinfo{journal}{\emph{Proceedings of the 2018 ACM Symposium on Principles of Distributed Computing}} (\bibinfo{year}{2018}).
\newblock


\bibitem[Hou et~al\mbox{.}(2017)]%
        {hou2017understanding}
\bibfield{author}{\bibinfo{person}{Binbing Hou}, \bibinfo{person}{Feng Chen}, \bibinfo{person}{Zhonghong Ou}, \bibinfo{person}{Ren Wang}, {and} \bibinfo{person}{Michael Mesnier}.} \bibinfo{year}{2017}\natexlab{}.
\newblock \showarticletitle{Understanding I/O performance behaviors of cloud storage from a client’s perspective}.
\newblock \bibinfo{journal}{\emph{ACM Transactions on Storage (TOS)}} \bibinfo{volume}{13}, \bibinfo{number}{2} (\bibinfo{year}{2017}), \bibinfo{pages}{1--36}.
\newblock


\bibitem[iEXEC(2017)]%
        {iexec}
\bibfield{author}{\bibinfo{person}{iEXEC}.} \bibinfo{year}{2017}\natexlab{}.
\newblock \bibinfo{title}{Blockchain-based decentralized cloud computing}.
\newblock \bibinfo{howpublished}{\url{https://iex.ec/wp-content/uploads/pdf/iExecWPv3.0-English.pdf}}.
\newblock


\bibitem[Initiative(2021)]%
        {openapi}
\bibfield{author}{\bibinfo{person}{OpenAPI Initiative}.} \bibinfo{year}{15 February 2021}\natexlab{}.
\newblock \bibinfo{title}{OpenAPI Specification v3.1.0}.
\newblock \bibinfo{howpublished}{\url{https://github.com/OAI/OpenAPI-Specification/}}.
\newblock


\bibitem[Jones et~al\mbox{.}(2015)]%
        {RFC7519}
\bibfield{author}{\bibinfo{person}{M. Jones}, \bibinfo{person}{J. Bradley}, {and} \bibinfo{person}{N. Sakimura}.} \bibinfo{year}{2015}\natexlab{}.
\newblock \bibinfo{booktitle}{\emph{JSON Web Token (JWT)}}.
\newblock \bibinfo{type}{RFC} 7519. \bibinfo{institution}{RFC Editor}.
\newblock
\showISSN{2070-1721}
\urldef\tempurl%
\url{http://www.rfc-editor.org/rfc/rfc7519.txt}
\showURL{%
\tempurl}
\newblock
\shownote{\url{http://www.rfc-editor.org/rfc/rfc7519.txt}}.


\bibitem[Jones and Hardt(2012)]%
        {RFC6750}
\bibfield{author}{\bibinfo{person}{M. Jones} {and} \bibinfo{person}{D. Hardt}.} \bibinfo{year}{2012}\natexlab{}.
\newblock \bibinfo{booktitle}{\emph{The OAuth 2.0 Authorization Framework: Bearer Token Usage}}.
\newblock \bibinfo{type}{RFC} 6750. \bibinfo{institution}{RFC Editor}.
\newblock
\showISSN{2070-1721}
\urldef\tempurl%
\url{http://www.rfc-editor.org/rfc/rfc6750.txt}
\showURL{%
\tempurl}
\newblock
\shownote{\url{http://www.rfc-editor.org/rfc/rfc6750.txt}}.


\bibitem[Kaleem and Shi(2021)]%
        {10.1007/978-3-662-63958-0_10}
\bibfield{author}{\bibinfo{person}{Mudabbir Kaleem} {and} \bibinfo{person}{Weidong Shi}.} \bibinfo{year}{2021}\natexlab{}.
\newblock \showarticletitle{Demystifying Pythia: A Survey of ChainLink Oracles Usage on Ethereum}. In \bibinfo{booktitle}{\emph{Financial Cryptography and Data Security. FC 2021 International Workshops}}, \bibfield{editor}{\bibinfo{person}{Matthew Bernhard}, \bibinfo{person}{Andrea Bracciali}, \bibinfo{person}{Lewis Gudgeon}, \bibinfo{person}{Thomas Haines}, \bibinfo{person}{Ariah Klages-Mundt}, \bibinfo{person}{Shin'ichiro Matsuo}, \bibinfo{person}{Daniel Perez}, \bibinfo{person}{Massimiliano Sala}, {and} \bibinfo{person}{Sam Werner}} (Eds.). \bibinfo{publisher}{Springer Berlin Heidelberg}, \bibinfo{address}{Berlin, Heidelberg}, \bibinfo{pages}{115--123}.
\newblock
\showISBNx{978-3-662-63958-0}


\bibitem[Knative({[n.\,d.]})]%
        {knative}
\bibfield{author}{\bibinfo{person}{Knative}.} \bibinfo{year}{[n.\,d.]}\natexlab{}.
\newblock \bibinfo{title}{Eventing}.
\newblock \bibinfo{howpublished}{\url{https://knative.dev/docs/eventing/}}.
\newblock
\newblock
\shownote{Last accessed on 5/28/2021}.


\bibitem[Kubeless({[n.\,d.]})]%
        {kubeless}
\bibfield{author}{\bibinfo{person}{Kubeless}.} \bibinfo{year}{[n.\,d.]}\natexlab{}.
\newblock \bibinfo{title}{Kubernetes native serverless framework}.
\newblock \bibinfo{howpublished}{\url{https://github.com/kubeless/kubeless}}.
\newblock
\newblock
\shownote{Accessed: 10/17/2022}.


\bibitem[Kursawe(2022)]%
        {beyondstaking}
\bibfield{author}{\bibinfo{person}{Klaus Kursawe}.} \bibinfo{year}{2022}\natexlab{}.
\newblock \showarticletitle{Beyond Staking: An Aphoristic design for Staking and Rewards}. In \bibinfo{booktitle}{\emph{The 2nd Workshop on Decentralized Finance (DeFi)}}.
\newblock


\bibitem[Li et~al\mbox{.}(2019)]%
        {li2019service}
\bibfield{author}{\bibinfo{person}{Wubin Li}, \bibinfo{person}{Yves Lemieux}, \bibinfo{person}{Jing Gao}, \bibinfo{person}{Zhuofeng Zhao}, {and} \bibinfo{person}{Yanbo Han}.} \bibinfo{year}{2019}\natexlab{}.
\newblock \showarticletitle{Service mesh: Challenges, state of the art, and future research opportunities}. In \bibinfo{booktitle}{\emph{2019 IEEE International Conference on Service-Oriented System Engineering (SOSE)}}. IEEE, \bibinfo{pages}{122--1225}.
\newblock


\bibitem[libp2p(2022a)]%
        {floodsub}
\bibfield{author}{\bibinfo{person}{libp2p}.} \bibinfo{year}{2022}\natexlab{a}.
\newblock \bibinfo{booktitle}{\emph{floodsub: Also known as pubsub-flood or just dumbsub, this implementation of pubsub focused on delivering an API for Publish/Subscribe, but with no CastTree Forming (it just floods the network).}}
\newblock
\urldef\tempurl%
\url{https://github.com/libp2p/js-libp2p-floodsub}
\showURL{%
\tempurl}


\bibitem[libp2p(2022b)]%
        {specspubsubatmasterlibp2pspecs-2022-12-16}
\bibfield{author}{\bibinfo{person}{libp2p}.} \bibinfo{year}{2022}\natexlab{b}.
\newblock \bibinfo{booktitle}{\emph{specs/pubsub at master · libp2p/specs}}.
\newblock
\urldef\tempurl%
\url{https://github.com/libp2p/specs/tree/master/pubsub}
\showURL{%
\tempurl}


\bibitem[libp2p(2022c)]%
        {WhatisPublishSubscribelibp2p-2022-12-16}
\bibfield{author}{\bibinfo{person}{libp2p}.} \bibinfo{year}{2022}\natexlab{c}.
\newblock \bibinfo{booktitle}{\emph{What is Publish/Subscribe - libp2p}}.
\newblock
\urldef\tempurl%
\url{https://docs.libp2p.io/concepts/pubsub/overview/}
\showURL{%
\tempurl}


\bibitem[Liu et~al\mbox{.}(2021)]%
        {9750531}
\bibfield{author}{\bibinfo{person}{Tao Liu}, \bibinfo{person}{Xiaowei Chen}, \bibinfo{person}{Jin Li}, \bibinfo{person}{Shaocheng Wu}, \bibinfo{person}{Wenlong Sun}, {and} \bibinfo{person}{Yueming Lu}.} \bibinfo{year}{2021}\natexlab{}.
\newblock \showarticletitle{Research on Progress of Blockchain Access Control}. In \bibinfo{booktitle}{\emph{2021 IEEE Sixth International Conference on Data Science in Cyberspace (DSC)}}. \bibinfo{pages}{516--522}.
\newblock
\urldef\tempurl%
\url{https://doi.org/10.1109/DSC53577.2021.00082}
\showDOI{\tempurl}


\bibitem[Maesa et~al\mbox{.}(2019)]%
        {accesscontrol}
\bibfield{author}{\bibinfo{person}{Damiano Di~Francesco Maesa}, \bibinfo{person}{Paolo Mori}, {and} \bibinfo{person}{Laura Ricci}.} \bibinfo{year}{2019}\natexlab{}.
\newblock \showarticletitle{A blockchain based approach for the definition of auditable Access Control systems}.
\newblock \bibinfo{journal}{\emph{Computers \& Security}}  \bibinfo{volume}{84} (\bibinfo{year}{2019}), \bibinfo{pages}{93--119}.
\newblock
\urldef\tempurl%
\url{https://doi.org/10.1016/J.COSE.2019.03.016}
\showDOI{\tempurl}


\bibitem[MetaMask(2022)]%
        {metamask}
\bibfield{author}{\bibinfo{person}{MetaMask}.} \bibinfo{year}{23 April 2022}\natexlab{}.
\newblock \bibinfo{title}{https://metamask.io/}.
\newblock \bibinfo{howpublished}{\url{https://metamask.io/}}.
\newblock


\bibitem[Mitzenmacher(2001a)]%
        {963420}
\bibfield{author}{\bibinfo{person}{M. Mitzenmacher}.} \bibinfo{year}{2001}\natexlab{a}.
\newblock \showarticletitle{The power of two choices in randomized load balancing}.
\newblock \bibinfo{journal}{\emph{IEEE Transactions on Parallel and Distributed Systems}} \bibinfo{volume}{12}, \bibinfo{number}{10} (\bibinfo{year}{2001}), \bibinfo{pages}{1094--1104}.
\newblock
\urldef\tempurl%
\url{https://doi.org/10.1109/71.963420}
\showDOI{\tempurl}


\bibitem[Mitzenmacher(2001b)]%
        {mitzenmacher2001power}
\bibfield{author}{\bibinfo{person}{Michael Mitzenmacher}.} \bibinfo{year}{2001}\natexlab{b}.
\newblock \showarticletitle{The power of two choices in randomized load balancing}.
\newblock \bibinfo{journal}{\emph{IEEE Transactions on Parallel and Distributed Systems}} \bibinfo{volume}{12}, \bibinfo{number}{10} (\bibinfo{year}{2001}), \bibinfo{pages}{1094--1104}.
\newblock


\bibitem[Network(2016)]%
        {golem}
\bibfield{author}{\bibinfo{person}{G. Network}.} \bibinfo{year}{2016}\natexlab{}.
\newblock \bibinfo{title}{Golem network: Online white paper}.
\newblock \bibinfo{howpublished}{\url{https://golem.network/doc/Golemwhitepaper.pdf}}.
\newblock


\bibitem[Nouman~Durrani and Shamsi(2014)]%
        {10.5555/2580126.2580611}
\bibfield{author}{\bibinfo{person}{Muhammad Nouman~Durrani} {and} \bibinfo{person}{Jawwad~A. Shamsi}.} \bibinfo{year}{2014}\natexlab{}.
\newblock \showarticletitle{Review: Volunteer Computing: Requirements, Challenges, and Solutions}.
\newblock \bibinfo{journal}{\emph{J. Netw. Comput. Appl.}}  \bibinfo{volume}{39} (\bibinfo{date}{mar} \bibinfo{year}{2014}), \bibinfo{pages}{369–380}.
\newblock
\showISSN{1084-8045}


\bibitem[OpenFaaS({[n.\,d.]})]%
        {openfaas}
\bibfield{author}{\bibinfo{person}{OpenFaaS}.} \bibinfo{year}{[n.\,d.]}\natexlab{}.
\newblock \bibinfo{title}{Openfaas - serverless functions made simple}.
\newblock \bibinfo{howpublished}{\url{https://github.com/openfaas/faas}}.
\newblock
\newblock
\shownote{Accessed: 09/11/2022}.


\bibitem[OpenWhisk({[n.\,d.]})]%
        {whisk}
\bibfield{author}{\bibinfo{person}{OpenWhisk}.} \bibinfo{year}{[n.\,d.]}\natexlab{}.
\newblock \bibinfo{title}{Apache openwhisk}.
\newblock \bibinfo{howpublished}{\url{https://github.com/apache/incubator-openwhisk}}.
\newblock
\newblock
\shownote{Accessed: 08/1/2020}.


\bibitem[Pillai et~al\mbox{.}(2022)]%
        {9756564}
\bibfield{author}{\bibinfo{person}{Babu Pillai}, \bibinfo{person}{Kamanashis Biswas}, \bibinfo{person}{Zhé Hóu}, {and} \bibinfo{person}{Vallipuram Muthukkumarasamy}.} \bibinfo{year}{2022}\natexlab{}.
\newblock \showarticletitle{Cross-Blockchain Technology: Integration Framework and Security Assumptions}.
\newblock \bibinfo{journal}{\emph{IEEE Access}}  \bibinfo{volume}{10} (\bibinfo{year}{2022}), \bibinfo{pages}{41239--41259}.
\newblock
\urldef\tempurl%
\url{https://doi.org/10.1109/ACCESS.2022.3167172}
\showDOI{\tempurl}


\bibitem[Robinson(2021)]%
        {ROBINSON2021108488}
\bibfield{author}{\bibinfo{person}{Peter Robinson}.} \bibinfo{year}{2021}\natexlab{}.
\newblock \showarticletitle{Survey of crosschain communications protocols}.
\newblock \bibinfo{journal}{\emph{Computer Networks}}  \bibinfo{volume}{200} (\bibinfo{year}{2021}), \bibinfo{pages}{108488}.
\newblock
\showISSN{1389-1286}
\urldef\tempurl%
\url{https://doi.org/10.1016/j.comnet.2021.108488}
\showDOI{\tempurl}


\bibitem[Sakimura et~al\mbox{.}(2014)]%
        {openid}
\bibfield{author}{\bibinfo{person}{N. Sakimura}, \bibinfo{person}{J. Bradley}, {and} \bibinfo{person}{M. Jones}.} \bibinfo{year}{2014}\natexlab{}.
\newblock \bibinfo{title}{OpenID Connect Dynamic Client Registration 1.0 incorporating errata set 1. OpenID Foundation}.
\newblock \bibinfo{howpublished}{\url{http://openid.net/specs/openid-connect-registration-1_0.html}}.
\newblock


\bibitem[Sheffer et~al\mbox{.}(2020)]%
        {RFC8725}
\bibfield{author}{\bibinfo{person}{Y. Sheffer}, \bibinfo{person}{D. Hardt}, {and} \bibinfo{person}{M. Jones}.} \bibinfo{year}{2020}\natexlab{}.
\newblock \bibinfo{booktitle}{\emph{JSON Web Token Best Current Practices}}.
\newblock \bibinfo{type}{BCP} 225. \bibinfo{institution}{RFC Editor}.
\newblock
\showISSN{2070-1721}


\bibitem[SONM(2017)]%
        {sonm}
\bibfield{author}{\bibinfo{person}{SONM}.} \bibinfo{year}{2017}\natexlab{}.
\newblock \bibinfo{title}{Supercomputer organized by network mining}.
\newblock \bibinfo{howpublished}{\url{https://whitepaper.io/document/326/sonm-whitepaper}}.
\newblock


\bibitem[TriggerMesh({[n.\,d.]})]%
        {triggermesh}
\bibfield{author}{\bibinfo{person}{TriggerMesh}.} \bibinfo{year}{[n.\,d.]}\natexlab{}.
\newblock \bibinfo{title}{EveryBridge}.
\newblock \bibinfo{howpublished}{\url{https://triggermesh.com/cloud_native_integration_platform/everybridge/}}.
\newblock
\newblock
\shownote{Last accessed on 9/28/2020}.


\bibitem[Vieira et~al\mbox{.}(2019)]%
        {Vieira2019DisOpenFaaSAD}
\bibfield{author}{\bibinfo{person}{Lucas Vieira}, \bibinfo{person}{Adbys Vasconcelos}, \bibinfo{person}{{\'I}talo Batista}, \bibinfo{person}{Rodolfo Silva}, {and} \bibinfo{person}{Francisco~Vilar Brasileiro}.} \bibinfo{year}{2019}\natexlab{}.
\newblock \showarticletitle{DisOpenFaaS: A Distributed Function-as-a-Service Platform}. In \bibinfo{booktitle}{\emph{SBRC Companion}}.
\newblock


\bibitem[Vyzovitis et~al\mbox{.}(2020a)]%
        {vyzovitis2020gossipsub}
\bibfield{author}{\bibinfo{person}{Dimitris Vyzovitis}, \bibinfo{person}{Yusef Napora}, \bibinfo{person}{Dirk McCormick}, \bibinfo{person}{David Dias}, {and} \bibinfo{person}{Yiannis Psaras}.} \bibinfo{year}{2020}\natexlab{a}.
\newblock \showarticletitle{GossipSub: Attack-resilient message propagation in the Filecoin and ETH2. 0 networks}.
\newblock \bibinfo{journal}{\emph{arXiv preprint arXiv:2007.02754}} (\bibinfo{year}{2020}).
\newblock


\bibitem[Vyzovitis et~al\mbox{.}(2020b)]%
        {DBLP:journals/corr/abs-2007-02754}
\bibfield{author}{\bibinfo{person}{Dimitris Vyzovitis}, \bibinfo{person}{Yusef Napora}, \bibinfo{person}{Dirk McCormick}, \bibinfo{person}{David Dias}, {and} \bibinfo{person}{Yiannis Psaras}.} \bibinfo{year}{2020}\natexlab{b}.
\newblock \showarticletitle{GossipSub: Attack-Resilient Message Propagation in the Filecoin and {ETH2.0} Networks}.
\newblock \bibinfo{journal}{\emph{CoRR}}  \bibinfo{volume}{abs/2007.02754} (\bibinfo{year}{2020}).
\newblock
\showeprint[arXiv]{2007.02754}
\urldef\tempurl%
\url{https://arxiv.org/abs/2007.02754}
\showURL{%
\tempurl}


\bibitem[Wu et~al\mbox{.}(2021)]%
        {9569798}
\bibfield{author}{\bibinfo{person}{Lei Wu}, \bibinfo{person}{Yki Kortesniemi}, \bibinfo{person}{Dmitrij Lagutin}, {and} \bibinfo{person}{Maryam Pahlevan}.} \bibinfo{year}{2021}\natexlab{}.
\newblock \showarticletitle{The Flexible Interledger Bridge Design}. In \bibinfo{booktitle}{\emph{2021 3rd Conference on Blockchain Research \& Applications for Innovative Networks and Services (BRAINS)}}. \bibinfo{pages}{69--72}.
\newblock
\urldef\tempurl%
\url{https://doi.org/10.1109/BRAINS52497.2021.9569798}
\showDOI{\tempurl}


\bibitem[Yang et~al\mbox{.}(2020)]%
        {9057456}
\bibfield{author}{\bibinfo{person}{Caixia Yang}, \bibinfo{person}{Liang Tan}, \bibinfo{person}{Na Shi}, \bibinfo{person}{Bolei Xu}, \bibinfo{person}{Yang Cao}, {and} \bibinfo{person}{Keping Yu}.} \bibinfo{year}{2020}\natexlab{}.
\newblock \showarticletitle{AuthPrivacyChain: A Blockchain-Based Access Control Framework With Privacy Protection in Cloud}.
\newblock \bibinfo{journal}{\emph{IEEE Access}}  \bibinfo{volume}{8} (\bibinfo{year}{2020}), \bibinfo{pages}{70604--70615}.
\newblock
\urldef\tempurl%
\url{https://doi.org/10.1109/ACCESS.2020.2985762}
\showDOI{\tempurl}


\bibitem[Zhao et~al\mbox{.}(2022)]%
        {https://doi.org/10.48550/arxiv.2209.09367}
\bibfield{author}{\bibinfo{person}{Haidong Zhao}, \bibinfo{person}{Zakaria Benomar}, \bibinfo{person}{Tobias Pfandzelter}, {and} \bibinfo{person}{Nikolaos Georgantas}.} \bibinfo{year}{2022}\natexlab{}.
\newblock \bibinfo{title}{Supporting Multi-Cloud in Serverless Computing}.
\newblock
\newblock
\urldef\tempurl%
\url{https://doi.org/10.48550/ARXIV.2209.09367}
\showDOI{\tempurl}


\end{thebibliography}

\end{document}